\documentclass[draft]{agujournal2019}
\usepackage{url} 
\usepackage{lineno}
\usepackage[inline]{trackchanges} 
\usepackage{soul}
\usepackage{longtable}
\usepackage[export]{adjustbox}
\usepackage{wrapfig}
\usepackage{amssymb}
\usepackage{multirow}
\usepackage{tablefootnote}
\usepackage{colortbl}

\draftfalse

\journalname{JGR: Space Physics}

\begin{document}

\title{Whistler Waves in the foot of Quasi-Perpendicular Super-Critical Shocks}

\authors{Ahmad Lalti\affil{1,2}, Yuri V. Khotyaintsev\affil{1}, Daniel B. Graham\affil{1}, Andris Vaivads\affil{3}, Konrad Steinvall\affil{1,2}, Christopher T. Russell\affil{4}}

\affiliation{1}{Swedish Institute of Space Physics, Uppsala, Sweden}
\affiliation{2}{Space and Plasma Physics, Department of Physics and Astronomy, Uppsala University, Uppsala, Sweden}
\affiliation{3}{Space and Plasma Physics, School of Electrical Engineering and Computer Science, KTH Royal Institute of Technology, Stockholm, Sweden}
\affiliation{4}{Department of Earth, Planetary and Space Sciences, University of California, Los Angeles, USA}

\correspondingauthor{Ahmad Lalti}{ahmadl@irfu.se}

\begin{keypoints}
    \item Using the magnetospheric multiscale (MMS) spacecraft we characterize whistler waves upstream of quasi-perpendicular super-critical shocks. 

    \item Shock-reflected ions are found to be in resonance with the observed whistlers, indicating their importance for the generation of the waves.

    \item Results from a dispersion solver agrees with the observations supporting the kinetic cross-field streaming instability as a possible source.
\end{keypoints}

 \begin{abstract}
Whistler waves are thought to play an essential role in the dynamics of collisionless shocks. We use the magnetospheric multiscale (MMS) spacecraft to study whistler waves around the lower hybrid frequency, upstream of 11 quasi-perpendicular super-critical shocks. We apply the 4-spacecraft timing method to unambiguously determine the wave vector $\mathbf{k}$ of whistler waves. We find that the waves are oblique to the background magnetic field with a wave-normal angle between $20^{\circ}$ and $42^{\circ}$, a wavelength around 100 km which is close to the ion inertial length. We also find that $\mathbf{k}$ is predominantly in the same plane as the magnetic field and the normal to the shock. By combining this precise knowledge of $\mathbf{k}$ with high-resolution measurements of the 3D ion velocity distribution we show that a reflected ion beam is in resonance with the waves, opening up the possibility for wave-particle interaction between the reflected ions and the observed whistlers. The linear stability analysis of a system mimicking the observed distribution, suggests that such a system can produce the observed waves.
\end{abstract}

\section*{Plain Language Summary}

The interaction between waves and particles is proposed to be one of the main mechanisms for energy dissipation at collisionless plasma shock waves. Of particular interest are a type of waves called whistlers, they fall in a frequency range that allows interactions with both electrons and ions, making them important for energy transfer between the two species. Their mechanism of generation is still not fully understood. We use data from the 4 magnetospheric multiscale (MMS) spacecraft to unambiguously characterize whistler precursor waves at quasi-perpendicular super-critical shocks. We find that the  waves are oblique to the background magnetic field with wavelength around the width of the shock and a frequency around the lower hybrid frequency. We also find that the shock-reflected ions are at a velocity that allows them to exchange energy with the waves, making them a likely source. We confirm this conclusion by using a computer code to model the system and study its stability.

\section{Introduction}

Collisionless shocks, despite their ubiquity in astrophysical plasmas \cite{treumann2009,bykov2011}, are not yet fully understood, in particular which mechanisms provide plasma thermalization in the absence of collisions. One of the main mechanisms proposed is wave-particle interactions. Whistler waves are known to play an integral part in the dynamics and evolution of collisionless shocks. Their spatial and temporal scale  allows them to mediate energy between ions and electrons, paving the way for the thermalization of cold solar wind plasma as it passes the shock.

Whistler waves were first observed close to the earth's bow shock by the OGO 5 spacecraft  \cite{heppner1967ogo}. Later they were observed at planetary bow shocks of Venus, Mercury and Saturn (\citeA{russell2007upstream} and references therein), and at interplanetary shocks \cite{wilson2012observations}. The waves can be separated into different categories according to their frequency. The low-frequency whistlers $f \sim 10^{-2}$ Hz \citeA{fairfield1969} are connected to the shock and generated locally in the ion foreshock, by the ion-ion two stream instability. The high frequency whistlers $f \sim 10^2$ Hz are observed near the foot and the ramp of the shock \cite{hull2012,tokar1984} and are considered to be generated by the whistler anisotropy instability. At intermediate frequencies, are whistlers around the lower hybrid frequency $f \sim f_{LH} \sim 10^0 - 10^1$ Hz, first observed by \citeA{fairfield1974whistler}. The  generation mechanism of these waves is still under investigation. For super-critical quasi-perpendicular shocks three possible mechanisms have been proposed: internal shock generation \cite{fairfield1974whistler,krasnosel1991nature,sundkvist2012dispersive,dimmock2013dispersion}, generation by shock macro-dynamics and non-stationarity \cite{1989Galeev,krasnosel1985nonlinear,balikhin1997experimental}, and generation by ion or electron microinstabilities in the foot/ramp region \cite{orlowski1995damping,krauss1995electron}. One particular instability of interest is the kinetic cross-field streaming instability (KCFSI) between the reflected ion beam and the incoming solar wind  \cite{wu1983kinetic} (also known as the modified two stream instability). Several observational \cite{hoppe1981upstream,wilson2012observations,dimmock2013dispersion} and simulation \cite{hellinger1996whistler,muschietti2017two} studies suggested that this instability is responsible for the generation of the intermediate-frequency whistlers. 

To investigate the source of the observed whistlers it is necessary to characterize the wave properties, i.e. their frequency, phase velocity $\left(\mathbf{V}_{phase}\right)$ and polarization in the plasma frame, as well as the corresponding particle distributions. Using single-spacecraft measurements one can determine the wave normal direction using minimum variance analysis (MVA) of the magnetic field, but with a $\pi$ ambiguity in direction \cite{sonnerup1988minimum}. The earliest multi-spacecraft measurement was done by \citeA{russell1988multipoint}. They used 2 spacecraft to characterize two types of waves, precursor waves attached to the shock and upstream turbulence waves, upstream of both bow shocks and interplanetary (IP) shocks.  \citeA{balikhin1997experimental} and \citeA{dimmock2013dispersion} used 2 spacecraft to study whistlers upstream of super-critical quasi-perpendicular shocks. By timing the difference in the measured signal between the 2 spacecraft and using MVA they were able to get $\mathbf{V}_{phase}$, and hence determine the wave characteristics in the solar wind frame. Overall, the wave normals and spectra of the waves have been characterized well, but getting $\mathbf{V}_{phase}$ in the plasma frame was challenging, its determination was restricted to rare occurrences where spacecraft separation was adequate. Furthermore, particle distributions were typically unresolved. The question of the source of the intermediate-frequency $f \sim f_{LH}$ whistler waves at quasi-perpendicular super-critical shocks remains open.

The magnetospheric multiscale (MMS) spacecraft \cite{MMS_overview} is a constellation of 4 spacecraft in a tetrahedral formation equipped with high-resolution field and particle instruments that allows the exploration of the microphysics of collisionless shocks in an unprecedented way. \citeA{hull2020mms} used MMS data to conduct a case study about the generation mechanism and energetics of whistler waves upstream of quasi-perpendicular and super-critical shocks, with Alfv\'enic Mach number $M_a$ much larger than the nonlinear whistler critical Mach number $M_{cwn}$, so that no precursor whistlers are to be found and non-stationary behaviour of the shock is expected \cite{krasnoselskikh2002nonstationarity}. They found that the KCFSI of the reflected ions with the solar wind plasma is the likely source of the observed whistlers. Here we extend this study to a different regime of shocks with $M_a \lesssim M_{cwn}$.

\section{Observation}
 
 We analyze whistler precursors at 11 super-critical and quasi-perpendicular shocks, with $M_a$ ranging between 3.5 and 9.8 and $\theta_{Bn}$ between $55^{\circ}$ and $82^{\circ}$ (Table \ref{T1}).  For the magnetic field measurement we use the fluxgate magnetometer (FGM) \cite{russell2016} with sampling up to 128 Hz. The fast plasma investigation (FPI) \cite{pollock2016fast} measures the 3D distribution functions of electrons and ions with a cadence of 30 ms and 150 ms respectively and with an energy range from 10 eV to 30 keV.

To illustrate our whistler precursor characterization, we use the shock observed by MMS on 2017 November 24, shown in  Figure \ref{fig:fig1}. This shock has $\theta_{Bn} = 82^{\circ}$ and $M_a = 4.2$. All vectorial quantities are in the $\left(n,t_1,t_2\right)$ coordinate system, with $n$ being the normal to the shock determined using coplanarity theorem with the jumps in the velocity and the magnetic field, mixed mode 3 method given by eq. 10.17 in \citeA{schwartz1998}, $\hat{t_2}= \hat{n} \times \hat{B}$ and $\hat{t_1} = \hat{t_2} \times \hat{n}$. Furthermore, to make sure that the first quadrant in the $n-t_1$ plane contains the upstream magnetic field, we flip the sign of  $\hat{t_1}$ and  $\hat{t_2}$ whenever the upstream magnetic field has a negative $\hat{n}$ component.
Between 23:20:16 and 23:20:22 UT we see a slight increase in the magnetic field and density (panels a-b) and a slight decrease in the ion velocity (panel c). Concurrently, a reflected ion component (a component with positive normal velocity) is observed (panel d), and is delimited by the vertical dashed lines in panels b-e. All of these are signatures of a localized foot region. This is followed by the shock ramp identified by an abrupt increase in the magnetic field and ion and electron densities, and an abrupt decrease in the ion velocity between 23:20:22 and 23:20:25 UT. For this event $M_{a}$ is comparable to $M_{cwn}$ given the uncertainties (Table 1).

\subsection{Wave characterization}

Between 23:19:54 and 23:20:22 UT, upstream precursor waves are clearly visible in panel a.
 A wavelet power spectrum of the magnetic fluctuations is shown in Figure \ref{fig:fig1} (e). The waves have frequency slightly below $f_{LH}$, with the highest intensity near the edge of the shock foot and as we go upstream the intensity of the waves decreases. Using singular value decomposition (SVD \cite{santolik2003, taubenschuss2014wave}) we calculate the degree of polarization (panel f), planarity (panel g) and ellipticity (panel f) of the waves. We see that these waves are highly polarized with predominant planar polarization as is evident from panels g-h, this indicates that the ellipticity calculation is reliable. From panel (f) it is evident that the waves have circular right-hand polarization. Noting that the polarization in the plasma frame (discussed later) is also right handed, we conclude that these are whistler waves.
 
 Using 4-spacecraft measurements we can unambiguously determine the wave vector \textbf{k}. We apply a running wavelet transform to the \textbf{B} measurement at each of the spacecraft for the interval shaded in Figure \ref{fig:fig2} (a), giving a power spectrum evolution with time. Then in wavelet space, at each frequency and time step, phase shifts, $\Delta \Phi \left(\omega, t \right)$, between the signals are calculated. The separation between the spacecraft (Figure \ref{fig:fig2} (f) ) $\sim$15 km is much smaller than the wavelength of the waves, hence no spatial aliasing is expected. 
  Knowing the distance between the spacecraft $\left( \mathbf{\Delta R} \right)$ and taking advantage of its tetrahedral formation, we calculate \textbf{k} using 
  \begin{equation}
      \Delta \Phi \left(\omega,t\right) = \mathbf{k} \left(\omega,t \right) \cdot \Delta \mathbf{R}.
  \end{equation}
Figure \ref{fig:fig2} (b) shows the wave vector obtained by a weighted averaging of  $\mathbf{k} \left(\omega,t \right)$ over the frequencies at each time step and using the wavelet power as the weight. A weighted standard deviation provides an error on the measurement.
Averaging \textbf{k} over the shaded time interval and normalizing we get $\langle \hat{k} \rangle = \langle \mathbf{k} \rangle / \langle k \rangle  = \left(0.76, 0.65, -0.05\right) \pm \left(0.15, 0.05, 0.08\right)$. This vector is mostly in the coplanarity, $n-t_1$ plane. Figure \ref{fig:fig2} (g) shows a schematic of the spacecraft position, $\langle \hat{k} \rangle$ and upstream \textbf{B} projected on the ecliptic plane.

Figures \ref{fig:fig2} (c-e) show $\theta_{kB}$, the angle between  $\langle \textbf{k} \rangle$ and \textbf{B}, the wavelength and the phase speed in the spacecraft frame. Averaging these quantities over the same time interval we get $\theta_{kB} = 42^{\circ} \pm 6^{\circ}$ showing that the whistlers are oblique. We find an average $\lambda$ of $70 \pm 6$ km, which is comparable to the ion inertial length $d_i =  85$ km. The average phase speed in the spacecraft frame is $V_{p,sc} = 340 \pm 70$ km s$^{-1}$. To go to the plasma frame we use $\mathbf{V}_{p,pf} =\mathbf{V}_{p,sc} -\left( \mathbf{V} \cdot \langle \hat{k} \rangle\right) \langle \hat{k}\rangle$, where $\mathbf{V}$ is the relative velocity between frames, here taken to be the measured ion velocity averaged over the same interval $\mathbf{V}_i = \left(-293,250,28\right) \pm \left(30,10,30\right)$ km s$^{-1}$.  $\langle \hat{k} \rangle$ is almost perpendicular to $\mathbf{V}_i$, making the dot product in the above equation small. The resultant $\mathbf{V}_{p,pf}$ has a magnitude of 400 km s$^{-1}$, and has a positive normal component, showing that these waves propagate upstream. Finally, using $f_{pf} = f_{sc} - \mathbf{V} \cdot \langle \mathbf{k} \rangle/2\pi$, we find the average frequency in the plasma frame $f_{pf} $, to be $5.9 \pm 1.1 Hz = \left(0.95 \pm 0.23\right) f_{LH} $, where $f_{sc}$ is the spacecraft-frame frequency.

Applying the same analysis to all 11 events (Table \ref{T1}) we find that the precursor waves always propagate upstream at an oblique $\theta_{kB}$ that varies between $20^{\circ}$ and $42^{\circ}$, with a wavelength ranging from $0.7$ to $1.7 d_i$, and the plasma-frame frequency ranging from $0.3$ to $1.2 f_{LH}$. Two of the events had left-hand polarized waves in the spacecraft frame, which flips to right-hand polarization in the plasma frame consistent with the other events.  The wave vectors are in the $n-t_1$ plane within $\pm 20^{\circ}$. All of the wave vectors are in the first quadrant (positive $k_n$ and $k_{t1}$), i.e. pointing upstream of the shock. Knowing that the group velocity for whistler waves always lies in between $\mathbf{B}$ and $\mathbf{k}$, we conclude that it is also pointing in the upstream direction.

\subsection{Ion velocity distribution}

The foot region of quasi-perpendicular super-critical shocks contain 3 primary plasma components: the incoming solar wind and reflected  ions, and the electrons. Such a system can be unstable to a KCFSI \cite{wu1983kinetic}. In what follows we show that the most likely source of the observed whistlers is the instability generated by the relative drift between the reflected ion beam and the solar wind plasma.

For the system under investigation the scale length of the foot is of the order of the ion gyroradius, at these scales the ions are unmagnetized. In that case, when the reflected ion beam satisfies the resonance condition:
\begin{equation}
\label{equ1}
    V_{phase} = \mathbf{V}_{beam} \cdot \hat{k},
\end{equation}
the plasma and the waves will be able to exchange energy. Furthermore, for the waves to grow the slope of the ion velocity distribution function (VDF) in the wave vector direction should be positive, $\frac{\partial f}{\partial v_{k}} >0 $.
To verify whether the resonance condition (eq. \ref{equ1}) is satisfied by the observed VDFs we show in Figure \ref{fig:fig3} the 2D ion VDF reduced in the $k-t_2$ plane and  1D ion VDF reduced in the $\mathbf{k}$ direction for 3 events with different $\theta_{Bn}$. The VDFs are plotted in the electron rest frame. Overlaid on top of the VDF, in the shaded area, is the measured phase speed along with its $2\sigma$ interval. One can see from the 2D VDFs that the reflected beam is in resonance with the waves and the part of the VDF with $\frac{\partial f}{\partial v_{k}} >0 $ is within the resonant interval. The resonance condition is satisfied for all 11 events analyzed, which suggests that reflected ions generate the observed waves.

\section{Discussion}

To verify that the observed VDFs are unstable and can generate waves with the observed properties we use a simplified model where the three plasma components, i.e. the electrons and the incoming and reflected ions, are represented by Maxwellian distributions and study the wave growth using
the Bo kinetic dispersion solver \cite{huasheng2019unified}. Older dispersion solvers, like WHAMP \cite{ronnmark1982whamp}, do not take into consideration cross-field drifts. On the other hand, Bo solves for the roots of the magnetized kinetic dispersion relation allowing for cross-field drifts, opening up the possibility to explore the stability of more complex plasma configurations than what was allowed by older solvers.

\begin{figure*}
    \centering
    \includegraphics[scale=0.5, center]{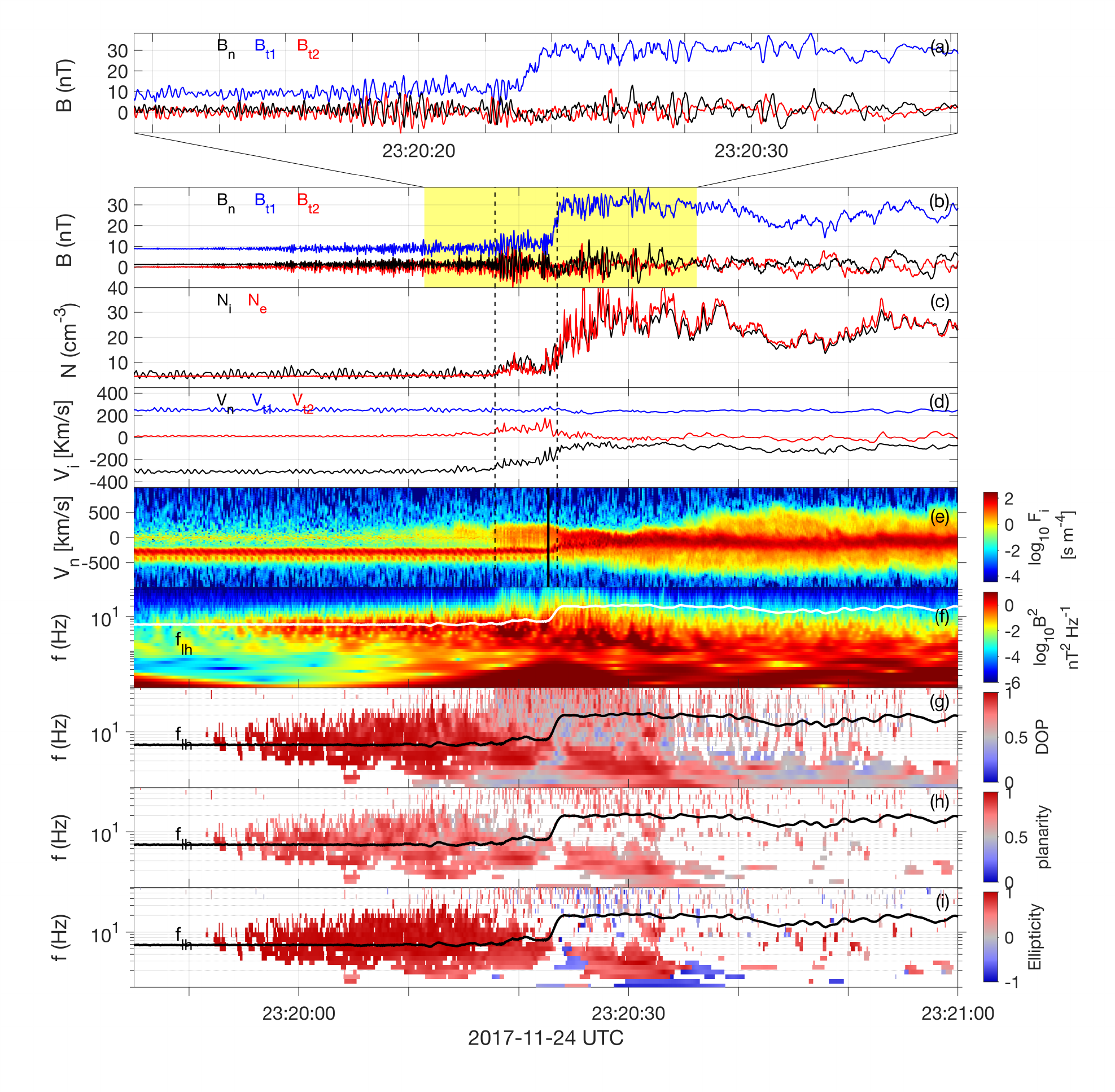}
    \caption{Overview of a shock crossing by MMS on 2017-11-24, 23:20 UT and wave polarization analysis. (a) Magnetic field enlarged around the foot and the ramp, (b) Magnetic field (c) electron and ion densities, (d) ion velocity, (e) 1D velocity distribution function reduced in the shock normal direction (f) power spectrum of the magnetic field, (g) degree of polarization, (h) planarity, and (i) ellipticity. Overlayed on top of panels (f-i) is the lower hybrid frequency. The black solid line in panel (e) shows the time of the 2D distribution of Figure \ref{fig:fig3} (a) corresponding to a peak in the current. The dashed lines in panels (b-e) mark the edges of the region with reflected ion component in the ion VDF}
    \label{fig:fig1}
\end{figure*}

\begin{figure*}
    \centering
    \includegraphics[scale=0.35, center]{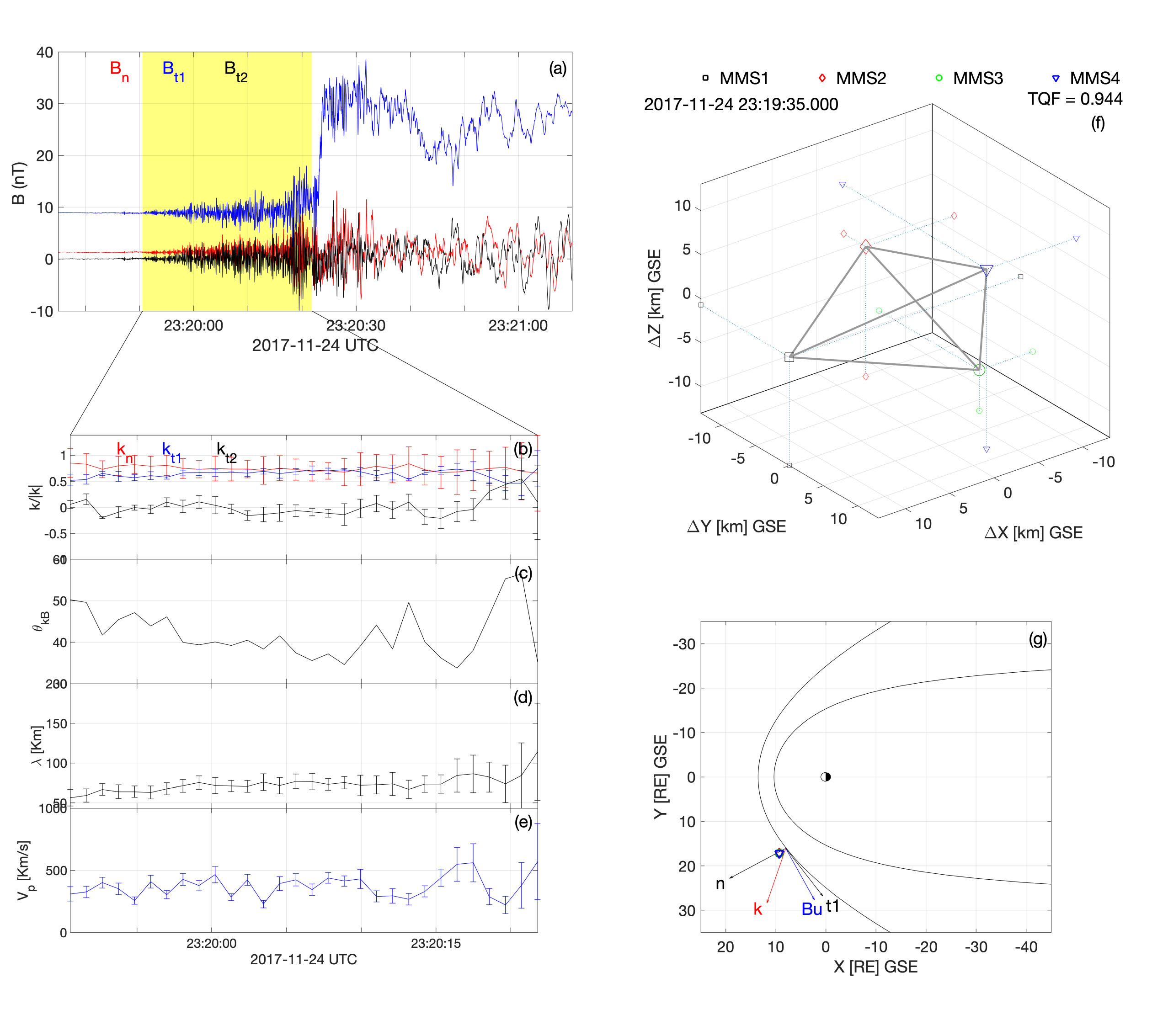}
    \caption{Wave characteristics. (a) Magnetic field $\mathbf{B}$, (b) wave vector $\mathbf{k}$, (c) the angle between $\mathbf{k}$ and the upstream $\mathbf{B}$, (d) wavelength, (e) phase velocity, (f) spacecraft configuration and (g) position of the spacecraft in the ecliptic plane with an averaged wave vector and upstream $\mathbf{B}$.}
    \label{fig:fig2}
\end{figure*}

\begin{figure*}
    \centering
    \includegraphics[scale=0.4, center]{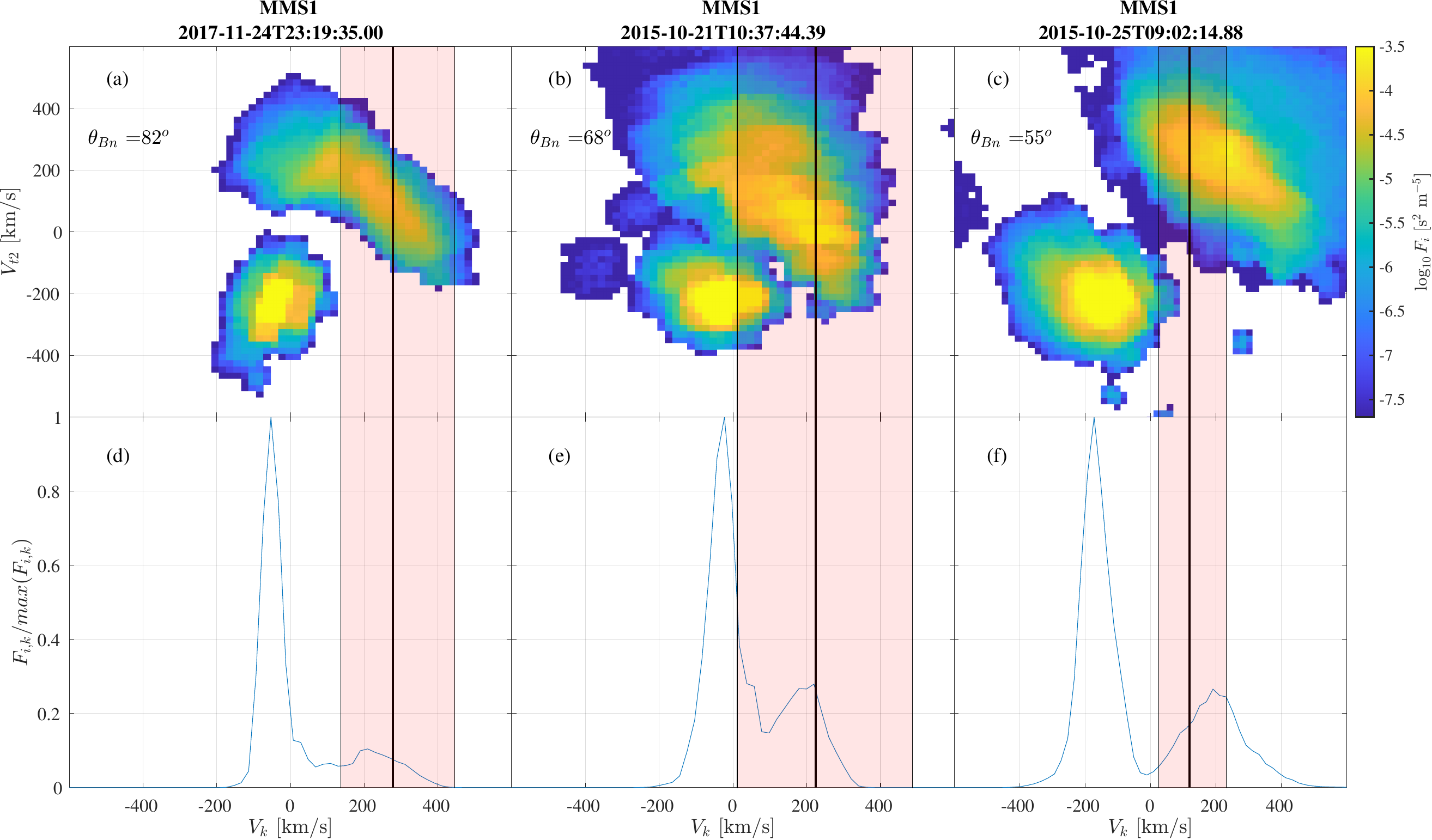}
    \caption{2D Ion VDFs reduced in the $k-t_2$ plane (top row) and 1D VDFs reduced in the k direction (bottom row)  for three different events with $\theta_{Bn} = 82^{\circ}$ (a,d), $68^{\circ}$ (b,e) and  $55^{\circ}$ (c,f). The black line shows the wave phase speed   and the pink shaded area shows its $2\sigma$ interval. The times indicate the center of the 150 ms acquisition interval of the ions VDFs.}
    \label{fig:fig3}
\end{figure*}

\begin{figure*}
    \centering
    \includegraphics[scale=0.35, center]{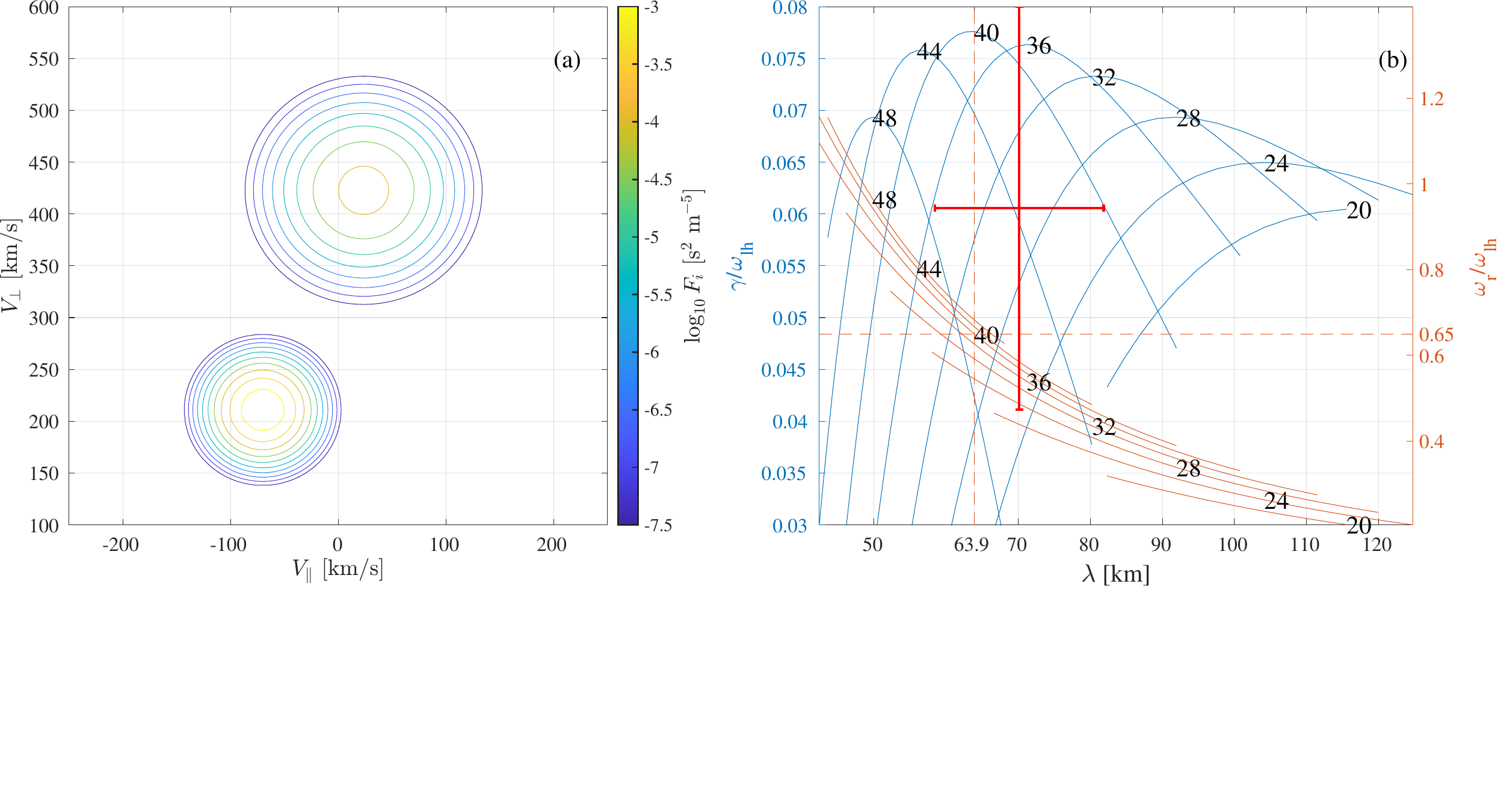}
    \caption{(a) The input ion VDF to the kinetic solver consisting of the incoming solar wind and reflected beam components. (b) Results of the kinetic solver: growth rate $\gamma$ and frequency $\omega_r$ normalized by $f_{LH}$ versus the wavelength $\lambda$ at different $\theta_{kB}$. The red error bars represent the measured $\lambda$ (horizontal) and  $\omega_r$ (vertical) with the $2 \sigma$ interval. The dashed lines show the values at maximum growth rate from the model.  The input parameters used for the solver are: $B= 9.3$ nT, solar wind ion density $n_{is} = 10.8$ cm$^{-3}$, and temperature $T_{is} = 2.5$ eV, reflected ion beam density $n_{ir} = 2.1$ cm$^{-3}$, and temperature $T_{ir} = 7.5$ eV, electron density $n_e = 12.9$ cm$^{-3}$, and temperature $T_{e} = 15.2$ eV. The solar wind and reflected ion velocities are $\mathbf{V}_{is} = \left(-70,-6 -210\right)$ km s$^{-1}$ and $\mathbf{V}_{ir} = \left(24,423, 0\right)$ km s$^{-1}$ with $\mathbf{V}_{e/SC-frame} = \left(-315, 360, 204\right)$ km s$^{-1}$ used to shift velocities from spacecraft to electron reference frame.}
    \label{fig:fig4}
\end{figure*}

We use measured values of B, density and temperature for the event shown in Figure \ref{fig:fig1} as input parameters for the solver. 
For ions we separate the VDF into an incoming (solar wind) and reflected beam components, and calculate their moments separately. We obtain the reflected and incoming ion density from the corresponding zeroth moment. The electron density is given by the sum of the two ion densities. As for the ion velocity we use first order moments of the VDF of each of the ion species. While for the electrons, FPI overestimates the velocity since it does not resolve low energy electrons, therefore we calculate the velocity from the current obtained by the curlometer method \cite{robert1998accuracy}, $\mathbf{V}_e = \mathbf{V}_i - \mathbf{J}/\left(ne\right)$. The velocities are transformed into a field-aligned coordinate system and to the rest frame of the electrons. For the temperature, we use FPI measurement for electrons. For ions, the solar wind beam is too narrow for FPI to properly characterise the temperature, so we use a time-shifted measurement from OMNI data. We treat the reflected ion beam temperature as a free parameter varied to give best agreement with the observed wave characteristics. Figure \ref{fig:fig4}a shows the modeled ion VDFs for the case of the shock event shown Figure \ref{fig:fig1}, which can be compared to the observed VDF in Figure \ref{fig:fig3}a. Summary of the input parameter is presented in Table \ref{T2}.

  \begin{table}[]
\renewcommand\thetable{2} 
    \centering
    \caption{Input Parameter to the Dispersion Solver}
    \label{T2}
\begin{longtable}{|l|c|c|c|}
\hline
& N (cm$^{-3}$)      & T (eV)               & \textbf{V}$\;^a$ (km s$^{-1}$)  \\
 \hline
Incoming ions       & 10.8 & 2.5    & -70 -6 -210 \\

Reflected ions      & 2.1  & 7.5    & 24 423 0    \\
Incoming electrons  & 12.9 & 15.2   & 0 0 0     \\
\hline
\multicolumn{3}{l}{\footnotesize{Background magnetic field used is B = 9.3 nT.}}\\
\multicolumn{3}{l}{\footnotesize{$^a$ All velocities are in electron reference frame. }}
\end{longtable}

\end{table}

The results of the solver obtained for a reflected beam temperature of $7.5$~eV (Figure \ref{fig:fig4}b) show that the system is unstable to whistler generation, with the maximum growth rate $\gamma_{max} \sim 0.1 \:\omega_{lh}=3$ (rad s$^{-1}$). However, for the waves to grow to large amplitudes they need to stay in the unstable region for a long enough time. We therefore calculate the spatial growth rate $\gamma_{s}=  \gamma/V_{gn}$, where $V_{gn}$ is the normal component of the group velocity in the shock reference frame. We then compare $\gamma_{s}^{-1}$ to the width of the foot $L_f = 0.68 V_{i}/\omega_{ci} = 0.68 \times 340/0.9 = 260$ km \cite{woods1971double} where $V_i$ is the upstream ion velocity in the shock frame. The product $\gamma_{s} L_f$ is the number of e-foldings of the wave while it propagates through the foot, which will then determine if the waves have sufficient time to grow to large amplitudes or not. For $\theta_{kB} = 40^\circ$, $\lambda = 64$ km and $f/f_{LH}=0.65$,  which are on the numerical dispersion surface and in closest agreement to the observed wave parameters, the model gives $ V_{gn} \sim 100$ km s$^{-1}$ with the spatial growth rate $\gamma_{s}= 0.03$ (rad km$^{-1}$) and $\gamma_s L_f \simeq 7.8 $ while the maximum $\gamma_s L_f \simeq 17 $. Thus, this linear model predicts that the waves can reach large amplitudes while propagating upstream within the foot region. This is consistent with the observation in Figure \ref{fig:fig1} (b-e) where the amplitude of the waves is small at the reflection point (rightmost dashed line in panels b-e), increases throughout the foot, and reaches a maximum near the upstream edge of the foot(leftmost dashed line in panels b-e). The results of this model suggest the large drift between the reflected ion beams and the incoming solar wind plasma as a possible driver of the observed whistlers.

Although the reflected beam is not a simple Maxwellian, approximating it by one can still qualitatively reproduce the physical behavior in terms of the wave properties and linear growth. Despite its complicated shape (see Figure \ref{fig:fig3}) the distribution is a monotonic function, and a Maxwellian captures qualitatively the gradients necessary for wave growth and damping. The dispersion solver only provides approximate values for the growth rate, wave vector and frequency as the phase-space gradients driving the waves are somewhat different from the observed gradients. Furthermore, we note that we have considered the linear growth in homogeneous plasma, and our model does not include the effects of inhomogeneity and non-linearity. These can be important if we intend to study growth and propagation of the large-amplitude waves in realistic shocks, and these effects can be addressed only in full kinetic simulations, which are outside of the scope of this paper. 

We note that \citeA{hull2020mms} studied whistler generation at a shock with $\theta_{Bn}=82^{\circ}$ and $M_a = 10$. They found that the reflected ion beam is in resonance with the waves, generating a whistler at frequency $\sim f_{lh}$ through the KCFSI. In their case $M_a$ exceeds the non-linear whistler critical Mach number, while our study is limited to shocks with $M_a \lesssim M_{cwn}$. Thus, the generation mechanism can be the same for shocks with Mach numbers both lower and larger than the non-linear whistler critical Mach number.

Apart from the instability discussed above, there are other mechanisms that can potentially generate oblique whistlers. For shock macro-dynamics\cite{1989Galeev,krasnosel1985nonlinear,balikhin1997experimental}, we study super-critical shocks with $M_a \lesssim M_{cwn}$, so no large scale dynamics is expected, hence we can rule out this mechanism as a source for the observed whistlers. Furthermore, other microinstabilities of ions can be active  in the foot/ramp region generating waves in the same frequency range, such as the lower hybrid drift instability, or the ion-ion drift instability  \cite{1984wu,1986scudder}. The lower hybrid drift instability predicts that the waves are oblique to the background magnetic field and are perpendicular to the coplanarity plane, while the ion-ion drift instability predicts waves that are perpendicular to the background magnetic field. Those predictions do not match the observed properties of the waves as is seen in Table \ref{T1}, hence those instabilities could not be behind the observed whistlers. As for instabilities involving electrons, they could be either due to  temperature anisotropy or loss-cone distributions or due to an electron beam \cite{hull2012,tokar1984}. The former generate whistlers that are predominantly parallel to the magnetic field which doesn't match the oblique whistlers observed. While to check for the generation by an electron beam we have calculated the energy necessary for the electron beam to be in resonance with the observed waves, and examining the electron VDF, we found no signature of a beam in the range of resonant energies, so such an instability could not be generating the observed whistlers. 

Finally, the whistlers can be generated by the shock itself and dispersively run upstream \cite{krasnoselskikh2002nonstationarity}. This model has two regimes separated by the linear whistler Mach number $M_w = |cos\left(\theta_{Bn}\right)| m_i /2 m_e$ where $m_e$ and $m_i$ are the electron and ion mass respectively \cite{krasnoselskikh2002nonstationarity}. When $M_a<M_w$ the wavelength of the waves is predicted to be around the ion skin depth, on the other hand when $M_a>M_w$ the wavelength of the waves is predicted to be around the electron skin depth. Looking at Table \ref{T1}, and taking into consideration the uncertainties, all events studied have $M_a \lesssim M_w$, so for them the predicted wavelength is similar to the observed one. Another key parameter that characterize the waves would be the wave vector direction. As it stands now, the dispersive shock model is a 1D model that assumes that the wave vector is parallel to the shock normal, which is clearly in contrast to what we observe (see Table \ref{T1}). Unfortunately, there is no 2D or 3D model of this mechanism available as of today, so we have no prediction for the wave normal direction that can be tested against observation in order to support or reject this mechanism. 



\section{Conclusions }
We use MMS spacecraft to study precursor whistler waves around the lower hybrid frequency at quasi-perpendicular supercritical shocks. We select 11 shock events with narrow band whistler precursor with an Alfv\'enic Mach number ranging between 3.5 and 9.8 (all $\lesssim M_{cwn}$) and $\theta_{Bn}$ between $55^{\circ}$ and $82^{\circ}$ for which the precursors can be characterized with high precision using the multi-spacecraft methods.

Our main findings are:
\begin{itemize}
    \item The wavelength of these waves range from 0.7 to 1.7 ion inertial length, the wave-normal angle range from $20^{\circ}$ to $42^{\circ}$ with a $\mathbf{k}$ directed upstream of the shock and close to the shock coplanarity plane. The frequency of the waves in the solar wind frame ranges between $0.3f_{LH}$ and $1.2f_{LH}$.
    \item The highest wave amplitude is found in the foot, where we found the shock-reflected ion component in the distribution function. After reducing the observed 3D ion VDF in the direction of $\mathbf{k}$ we find that the reflected ion component of the VDF is in Landau resonance with the observed waves, which indicates that the reflected ion beam is interacting with the observed whistlers and could be behind the generation of those waves.
    \item Using a linear kinetic dispersion solver we find that a VDF composed of a reflected ion beam on top of incoming solar wind, with parameter taken from observation, is unstable to generation of whistler waves with properties close to what we observe. This supports the kinetic cross-field streaming instability between the reflected ions beam and the incoming solar wind plasma as a likely generation mechanism.
\end{itemize}

We have found that the waves are in resonance with the shock reflected ions, and the linear stability analysis shows that the waves can potentially grow to large amplitudes driven by the drift between the reflected ions and the solar wind. But this analysis obviously does not include the effects of non-linearity and inhomogeneity which can affect the wave growth. Moreover, there is another possible generation mechanism (dispersive shock model) which predicts the waves with the observed wavelength. In order to determine which mechanism is generating the observed whistlers one has to compare the results of a full kinetic simulation to the observations. In this manuscript, we have presented the wave properties with high accuracy as well as presented the relevant observations of the ion VDFs which can be  used for such a comparison. 

 \acknowledgments
We thank the entire MMS team and instrument PIs for data access and support. We are grateful to Richard Denton for helping us with the Bo solver which can be found at \url{http://dx.doi.org/10.17632/cvbrftzfy5.1}. MMS data are available at \url{https://lasp.colorado.edu/mms/sdc/public/data/} following the directories: mms\#/fgm/brst/l2 for FGM data, mms\#/fpi/brst/l2/dis-dist for FPI ion distributions, mms\#/fpi/brst/l2/dis-moms for FPI ion moments and mms\#/fpi/brst/l2/des-moms for FPI electron moments. OMNI data used are available at \url{https://omniweb.gsfc.nasa.gov/}. Data analysis was performed using the IRFU-Matlab analysis package. No new data have been produced as part of this project. This work is supported by the Swedish Research Council grant 2018-05514.

\begin{sidewaystable}
\renewcommand\thetable{1} 
\caption{Shock and wave parameters for the analyzed events }
\label{T1}
\begin{adjustbox}{width=1\textwidth} 

\begin{tabular}{|c|c|c|c|c|c|c|c|c|c|c|c|c|c|c|}
\hline
\hline
Time  &  $\theta_{Bn}^{\circ}$   &     $k_{nt1t2}$    &   $Bu$ (nT)   &  $M_a$ \footnote{Uncertainties in Mach numbers are estimated from the uncertainty in shock normal determination}  &   $M_{w}^{\;\;\;\;\;\; a}$   &   $M_{cwn}^{\;\;\;\;\;\; a}$ &   $\lambda$ (km)    &   $Vp$ (km s$^{-1}$)     &   $\theta_{kb}^{o}$   &   $\theta_{kn}^{o}$   &  $f_{pl}$ (Hz)   &   $f_{pl}/f_{lh} $
\\ \hline
2017-11-24 23:19 & 82     & 0.76        0.65       -0.05 & 1.36        9.22           0   & 4.2 $\pm$ 0.1  & 3.0 $\pm$ 0.4 & 3.5 $\pm$ 0.6       & 70 $\pm$ 6          & 340 $\pm$ 70 & 42 $\pm$ 6 & 41 $\pm$ 7  &  5.9  $\pm$ 1.1   &     0.94 $\pm$ 0.23    \\
\hline
2019-05-15 15:45 & 61  & 0.8        0.51        0.32     & 3.62         7.1           0   & 5.4 $\pm$ 0.3  & 10.4 $\pm$ 0.3 & 12.8 $\pm$ 0.4    & 64 $\pm$ 9             & 100 $\pm$ 40  & 36 $\pm$ 6 & 37 $\pm$ 10        &  3.3 $\pm$  0.6  & 0.6 $\pm$ 0.18    \\
\hline
2017-10-18 04:32 & 57    &  0.92        0.4        -0.01  & 1.36        2.73           0   & 9 $\pm$ 1  & 12.2 $\pm$ 1.3 & 15 $\pm$ 2      & 125 $\pm$ 25        & 140 $\pm$ 30 & 40 $\pm$ 10  & 23 $\pm$ 14        &  1.06 $\pm$  0.25  &  0.51 $\pm$ 0.14   \\
\hline
2017-11-02 08:28 & 67   &  0.73       0.68           0   & -3.7        -8.91           0   & 4.1 $\pm$ 0.2  & 8.0 $\pm$ 0.8 & 10 $\pm$ 1          & 80 $\pm$ 9       & 181 $\pm$ 60  & 152 $\pm$ 6 & 43 $\pm$ 8       &   2.6 $\pm$  0.9 & 0.41 $\pm$ 0.15     \\
\hline
2018-02-21 09:40 & 60    & 0.74        0.64       -0.19   & 2.75        4.65           0   & 7.7 $\pm$ 0.3  & 11.5 $\pm$ 1.0 & 14 $\pm$ 1       & 75 $\pm$ 6         & 78 $\pm$ 18  & 23 $\pm$ 6  & 42 $\pm$ 5 &  3.1 $\pm$  0.5 &   0.86 $\pm$ 0.15     \\
\hline
2017-11-16 13:51 & 56   & 0.66        0.73        0.16   & 2.79        4.48           0   &7.4 $\pm$ 0.3  & 11.1 $\pm$  1.0 & 14 $\pm$ 1     & 121 $\pm$ 16      & 217 $\pm$ 50 & 20 $\pm$ 9  & 49 $\pm$ 5          &   1.5 $\pm$  0.5 &    0.35 $\pm$ 0.15      \\
\hline
2017-11-02 04:26 & 62   & 0.74       0.64       0.18   & -4.71        -9.43           0  & 3.5 $\pm$ 0.2  & 8.8 $\pm$ 1.0 & 11 $\pm$ 1     & 71 $\pm$ 16        & 80 $\pm$ 27  & 150 $\pm$ 10 & 42 $\pm$ 17       &   2.4  $\pm$ 0.5 &    0.33 $\pm$ 0.09     \\
\hline
2017-10-21 21:48 & 63   & 0.75       0.65        0.15   & 3.85        7.84           0   & 5.0 $\pm$ 0.5  & 9.1 $\pm$ 1.2 & 12 $\pm$ 2          & 110 $\pm$ 6       & 180 $\pm$ 40 & 30 $\pm$ 8  & 42 $\pm$ 9        &  2.3 $\pm$0.3    & 0.39 $\pm$ 0.07   \\
\hline
2015-10-21 10:37 & 68   & 0.78        0.63       -0.04   & 4.31        11.36            0 & 6.2 $\pm$ 0.2  & 6.3 $\pm$ 2 &  8 $\pm$ 3      & 40 $\pm$ 6         & 122 $\pm$ 67 & 34 $\pm$ 8 & 39 $\pm$ 13 &   8.8  $\pm$  1.7 &  1.1 $\pm$ 0.3     \\
\hline
2015-10-25 09:02 & 55    &  0.8        0.6        -0.05    & -3.73        -5.89           0  &  9.8 $\pm$ 0.3 & 11.6 $\pm$ 0.6 & 14.7 $\pm$ 0.9      & 44 $\pm$ 8        & 80 $\pm$ 40  & 157 $\pm$ 9  & 37 $\pm$ 14      &  6.1 $\pm$   1.0 &   1.2 $\pm$ 0.3     \\
\hline
2015-10-21 10:10 & 69 & 0.63        0.72         0.3   & 0.79        12.82            0 & 8.3 $\pm$ 0.4   & 8.8 $\pm$ 0.9 & 11 $\pm$ 1      & 39 $\pm$ 7       & 80 $\pm$ 30  & 40 $\pm$ 10  & 51 $\pm$ 13        &  4.7 $\pm$ 1.1    & 0.5 $\pm$ 0.21   \\
\hline
\end{tabular}
\end{adjustbox}
\end{sidewaystable}

\bibliography{biblio}

\end{document}